\documentstyle[aps,prl,epsfig,floats]{revtex}
\begin{document}
\twocolumn[
\hsize\textwidth\columnwidth\hsize\csname@twocolumnfalse\endcsname
\draft
\title{Experimental determination of $B$-$T$ phase diagram of
YBa$_{2}$Cu$_{3}$O$_{7 - \delta}$ to 150T for $B{\perp}c$}
\author{J. L. O'Brien,$^1$\cite{email} H. Nakagawa,$^2$ A. S. Dzurak,$^1$ R.
G. Clark,$^1$ B. E.  Kane,$^1$ N. E. Lumpkin,$^1$ N. Miura,$^2$ E. E.
Mitchell,$^3$ J. D. Goettee,$^4$ J. S. Brooks,$^5$ D. G. Rickel,$^4$ and R. P.
Starrett$^1$}   \address{$^1$National
Pulsed Magnet Laboratory and Semiconductor Nanofabrication Facility,\\
School
of Physics, University of New South Wales, Sydney 2052, Australia}
\address{$^2$Institute for Solid State Physics, University of Tokyo, 7-22-1
Roppongi, 106 Tokyo, Japan}
\address{$^3$CSIRO, Division of Telecommunications and Industrial Physics,
Lindfield 2070, Australia} 
\address{$^4$Los Alamos National Laboratory, Los Alamos,
New Mexico 87545} 
\address{$^5$Department of Physics, Florida State University, Tallahassee FL
32310}
\date{\today}
\maketitle 

\begin{abstract}
The $B$-$T$ phase diagram for thin film
YBa$_{2}$Cu$_{3}$O$_{7 - \delta}$ with $B$ parallel to the superconducting
layers has been constructed from GHz transport measurements to 150T. Evidence
for a transition from a high $T$ regime dominated by orbital effects, to a
low $T$ regime where paramagnetic limiting drives the quenching of
superconductivity, is seen. Up to 110T the upper critical field is found
to be linear in $T$ and in remarkable agreement with extrapolation of the
longstanding result of Welp {\em et al} arising from magnetisation
measurements to 6T. Beyond this a departure from linear behaviour occurs at
$T$=74K, where a 3D-2D crossover is expected to occur. \end{abstract} 

]
Recent magneto-transport measurements on high-$T_c$ cuprates have provided
invaluable information in building a complete picture of these materials,
essential to the development of a rigorous theory for the phenomenon of high
temperature superconductivity. For example, divergence in the upper critical
field at low $T$ has been reported in overdoped Tl$_{2}$Ba$_{2}$CuO$_{6}$
\cite{mackenzie}, Bi$_{2}$Sr$_{2}$CuO$_{y}$
\cite{osofsky,ando} and
La$_{2-x}$Sr$_{x}$CuO$_{4}$ \cite{ando}. Such results have provided the
impetus for considerations of unconventional behaviour in the anisotropic
high-$T_c$ cuprates such as reentrant superconductivity and the possibility to
exceed the paramagnetic limit \cite{lebed}. These measurements have typically
relied on millisecond pulsed fields to observe upper critical fields which
exceed the range ($\sim$35T) of steady field magnets. For
higher $T_c$ materials such as optimally oxygen-doped YBCO ($T_c{\sim}$90K)
access to the normal state requires magnetic fields well in excess of those
generated by ms pulsed systems except for $T$ very near $T_c$.
Explosive flux compression technology has been used to access this high field
regime \cite{dzurak,smith,goettee,bykov}, in one case providing evidence for
paramagnetic limiting of the upper critical field in YBCO for $B$ parallel
to the superconducting layers ($B{\perp}c$) \cite{dzurak}.
However, such measurements are extremely difficult and opportunities to make
them are few. Single-turn coil magnetic field generators have also
been used to make transport measurements on thin film YBCO \cite{nakagawa}.
While these systems do not produce fields in the range of flux compression
techniques, they do allow for systematic, repeatable measurements to be made
as the destruction of the coil does not damage the sample or cryostat.
Nevertheless, transport measurements remain difficult since the peak field is
reached in a few $\mu$s and the maximum d$B$/dt exceeds $10^8$T/s.  

Previous flux
compression measurements on thin film YBCO with $B{\perp}c$ \cite{dzurak} gave
an onset of dissipation $B_{ons}$=150T and an upper critical field
$B_{c2}$=240T at 1.6K. This $B_{c2}$ is significantly smaller than
$B_{c2}^{\perp}$(0)=674T predicted by Welp {\em et al} \cite{welp} who applied
the Werthamer-Helfand-Hohenberg (WHH) \cite{whh} formalism, accounting for
orbital effects only, following measurements of the slope d$B_{c2}$/d$T$ near
$T_c$. This large discrepancy has been interpreted \cite{dzurak} in terms of
the Clogston-Chandresekhar paramagnetic limit $B_p$ \cite{clogston,chandr}
which arises when the Zeeman energy exceeds the superconducting energy gap
${\Delta}_0$, thus destroying the Cooper pair singlet state. Within BCS theory
$B_{p}$=${\gamma}T_c$, with ${\gamma}$=1.84T/K \cite{clogston} which, for
optimally-doped YBCO gives $B_p{\sim}$170T. In stark contrast to these results
for in-plane magnetic fields, measurements \cite{smith,nakagawa} in the
alternative and more widely-studied configuration, $B{\parallel}c$, have mapped
out the entire phase diagram in good agreement with the WHH model.

Here we report transport
measurements which have allowed the $B$-$T$ phase diagram of YBCO to be
constructed for $B$$<$150T in the $B{\perp}c$ orientation, greatly extending
previous magnetisation measurements to 6T \cite{welp}. Although explosive flux
compression techniques have previously been used to access this
regime \cite{dzurak,goettee}, the single turn coil system allows
systematic measurements to be made on a single sample with both rising and
falling magnetic field. Measurements on thin-film samples were made using a GHz
technique \cite{kane} in a single-turn coil system \cite{miura} generating
fields to 150T. The superconducting-normal transition was observed to be an
equilibrium process, evidenced by the absence of any measurable hysteresis
between up and down $B$ sweeps. Above $T$=74K $B_{c2}$ is found to be linear
in $T$ with the slope $\alpha$=d$B_{c2}$/d$T$ corresponding closely to that
found in magnetisation measurements \cite{welp}. Below 74K a departure from
this slope is observed and is understood as arising from a transition from 3D
behaviour where orbital effects quench superconductivity to 2D behaviour where
paramagnetic limiting dominates.

\begin{figure}
\vspace{-2.5cm}
\begin{center}
\includegraphics[width=7.8cm,angle=270]{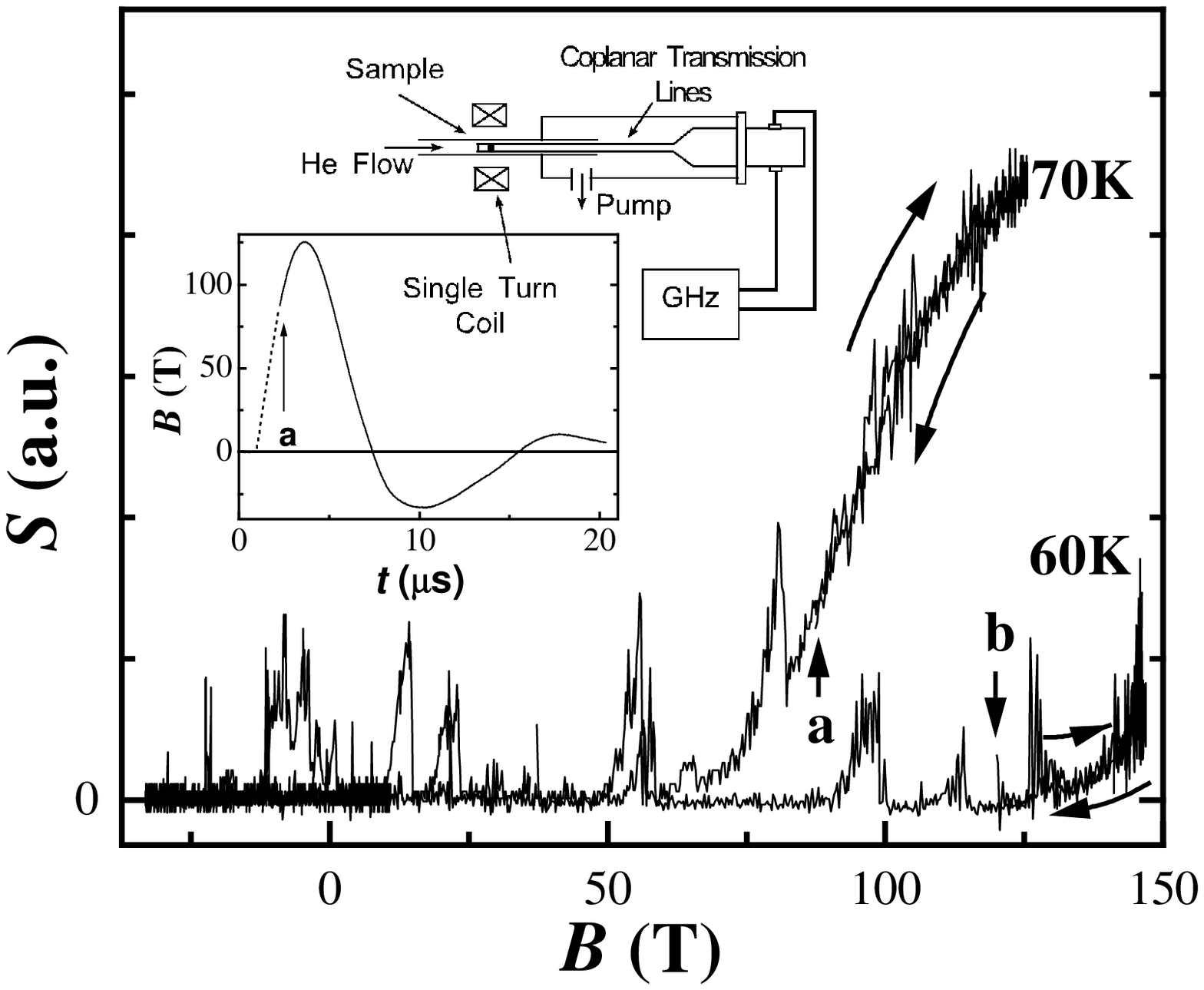}
\end{center}
\vspace{-.4cm}
\caption{Raw GHz transmission $S$, with $B{\perp}c$, shown for times after the
point marked {\bf a} ({\bf b}) for $T$=70K (60K). Before these
times, data are obscured by electrical noise arising from the discharge of the
capacitor bank. Arrows indicate the direction of the field sweep. The insets
show the experimental configuration for measurements and the field profile of
the single turn coil.}
\vspace{-.3cm}
\label{fig1}
\end{figure}

The
experimental configuration is shown in the inset to Fig.\ \ref{fig1}. A
symmetric triplet coplanar transmission line (CTL) carries a microwave
signal, ${\nu}$=0.8GHz-1GHz, past the sample and the transmission $S$ is
modulated by the resistivity $\rho$ of the sample. Full details of the
experimental arrangement are set out in Ref. \cite{kane}. A flow of cold
He gas gives access to $T$ in the range 7K-300K with sample $T$ monitored by a
AuFe-Cromel thermocouple mounted on the back side of the substrate. Two
thermocouples mounted side by side on the same sample gave consistent readings
to within 0.5K. Discharge of a 40kV capacitor bank into a 10mm diameter single
turn copper coil generated fields to 150T and the copper coil was vaporised in
the process, leaving the sample and cryostat intact \cite{miura}.  

A YBCO film, thickness 250nm, $T_c$=87.2K and critical current
$J_c$=3.14MA/cm$^2$ at 77K, was grown by on-axis dc magnetron sputtering
on a MgO (001) substrate, with the $c$-axis oriented in the growth direction,
at $T$=$770^{\circ}$C in an Argon-Oxygen atmosphere with a deposition time
${\sim}$90min. The film was etched to produce a 20${\mu}$m strip
perpendicular to the CTL to match the sample resistance to the
characteristic impedance of the CTL Z=50$\Omega$. A 50nm dielectric layer of
Si$_3$N$_4$ separated the film and CTL so that coupling to the sample was
capacitive. This eliminates the need for ohmic contacts to the sample which
can be problematic in pulsed fields \cite{lumpkin}.

Raw transmission $S$ as a function of $B$ is plotted for
$T$=60K and 70K in Fig.\ \ref{fig1}. The single-turn coil system produces a
number of cycles of $B$ prior to destruction (Fig.\ \ref{fig1} inset).
The absence of any hysteresis in the data was consistently observed in a number of samples at
a range of temperatures, providing confidence in the critical field
information obtained and in its comparison with equilibrium models for
high-$T_c$ behaviour.

\begin{figure}
\vspace{-2.5cm}
\begin{center}
\includegraphics[width=7.8cm,angle=270]{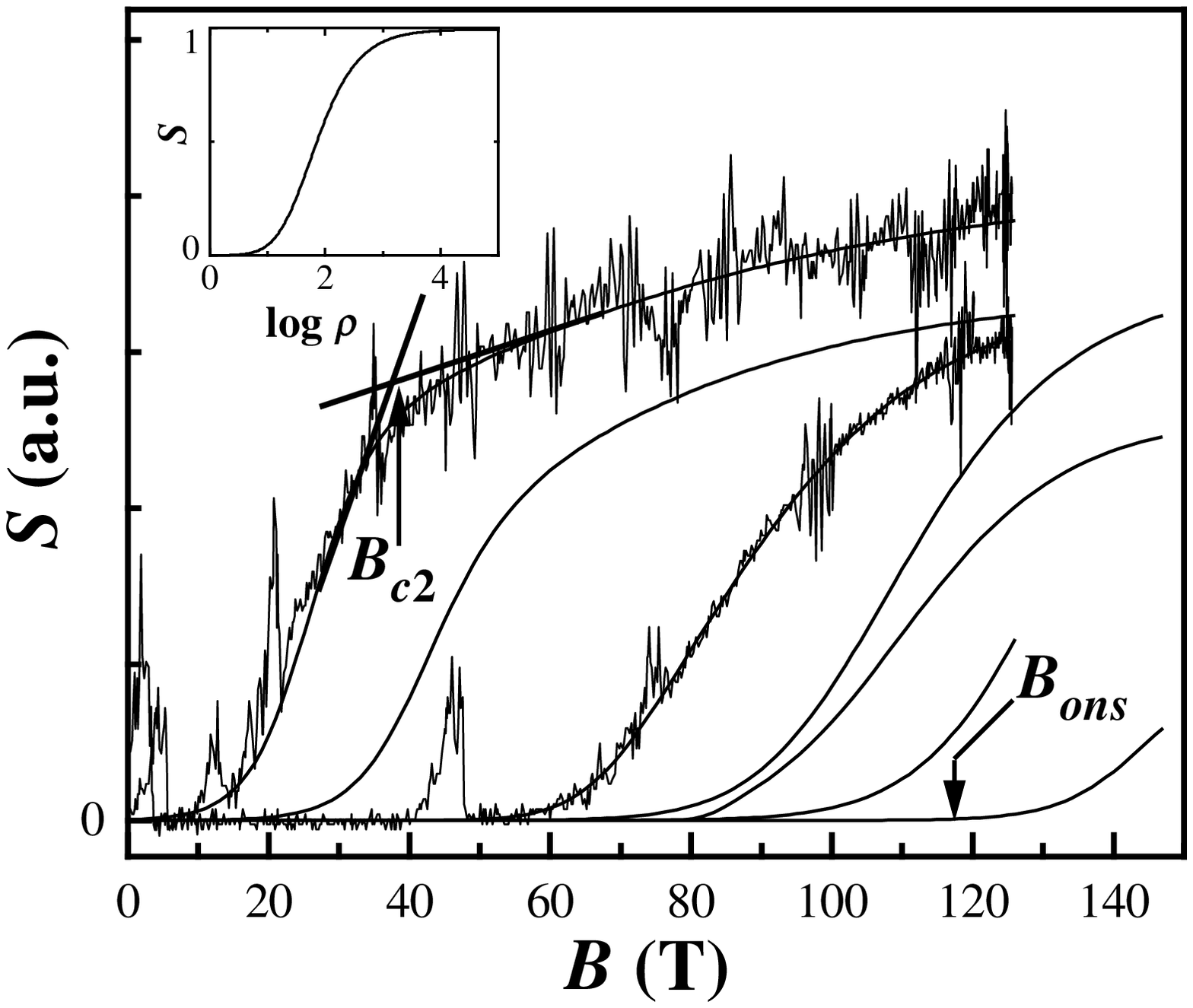}
\end{center}
\vspace{-0.4cm}
\caption{Fits to raw magneto-response $S(B)$ data for $T$=80K, 78K, 74K,
72K, 66K, 65K and 60K with onset fields increasing respectively. Traces have
been offset vertically for clarity. Raw data are shown for comparison for
$T$=80K and 74K. Definitions of $B_{ons}$ and $B_{c2}$ are indicated. The $S$
response to sample sheet resistivity $\rho$ (in $\Omega$) is shown in the
inset.}
\vspace{-.3cm}
\label{fig2}
\end{figure}

In the
superconducting state the sample acts as an equipotential across the
CTL, completely attenuating the microwave signal, resulting in
zero transmission. As the applied field $B$ drives the sample normal, $S$
increases with increasing $\rho$. A model of this response is shown in the
inset to Fig.\ \ref{fig2}, calculated assuming capacitive coupling across a
thin dielectric layer to a 2D sheet of electrons \cite{kane}. Sharp noise
spikes in the data are attributed to GHz emission from the plasma produced in
vaporisation of the single turn coil and are predominantly in the direction of
increasing $S$ since the technique measures transmitted power (as
opposed to voltage).

Fits to the raw data for the decreasing $B$ sweep which take into
account the positive-going nature of the noise spikes are shown in Fig. 2. We
define $B_{c2}$ as the intersection of the tangent to the transmission curve
in the transition region and that immediately after it, following Ref.
\cite{mackenzie} on the basis that this gives values close to the 90${\%}$
criterion and in good agreement with tunnelling data. For the lowest $T$
measurements, determining $B_{c2}$ becomes more difficult because there is
less saturation region. We find, however, that the form of the transition is
essentially the same for all $T$. $B_{ons}$ is defined as the point at which
the transmission departs from $S$=0.

\begin{figure}
\vspace{-2.5cm}
\begin{center}
\includegraphics[width=7.8cm,angle=270]{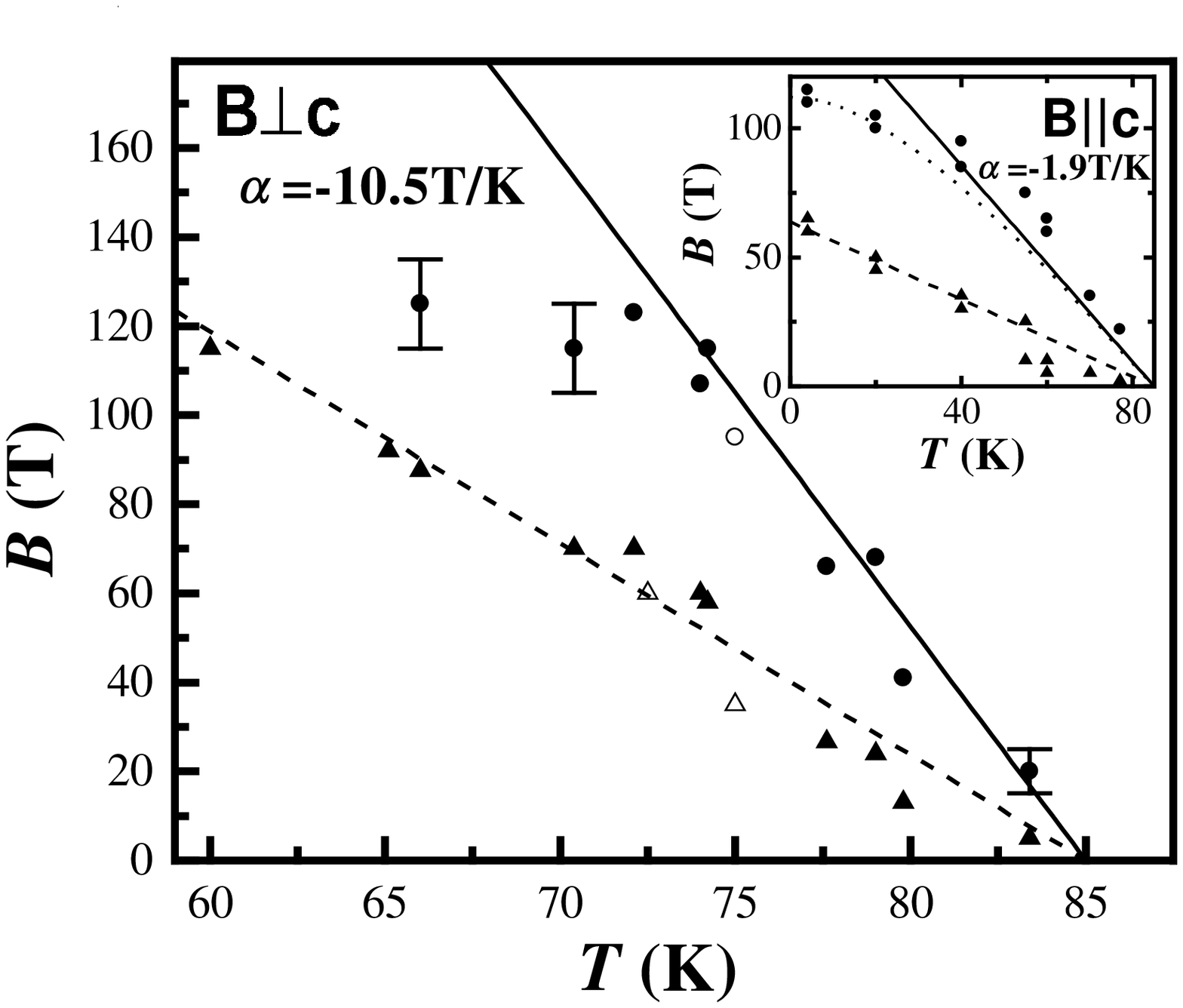}
\end{center}
\vspace{-.4cm}
\caption{$B$-$T$ phase diagram for YBCO with $B{\perp}c$. Solid circles
and triangles, represent $B_{c2}$ and $B_{ons}$ respectively. Open
symbols arise from measurements on a second sample fabricated from the same
film. The inset shows the equivalent phase diagram taken from Nakagawa {\em et
al} \protect\cite{nakagawa} with $B{\parallel}c$. In both cases the solid line
shows the slope $\alpha$=d$B_{c2}$/d$T$ determined from magnetisation
measurements \protect\cite{welp} and the dashed line is simply a guide to the
eye.} \vspace{-.3cm}
\label{fig3} 
\end{figure}

Fig.\ \ref{fig3} shows the $B$-$T$ phase diagram for $T$$>$60K, with $B_{ons}$
and $B_{c2}$ values determined from the complete data set, a subset of which is
shown in Fig.\ \ref{fig2}. In magnetisation measurements up to 6T on single
crystal YBCO with $B{\perp}c$, $B_{c2}$ was found to be linear in $T$ with
d$B_{c2}$/d$T$=-10.5T/K \cite{welp}. The extrapolation of this slope $\alpha$
is plotted in Fig.\ \ref{fig3} and we find that our data follow it closely
down to $T$=74K where $B_{c2}{\sim}$100T. Note that as in Ref. \cite{welp}
this line intersects the $T$ axis slightly below $T_c$ as do the $B_{ons}$
data. This is allowed for in Fig.\ \ref{fig4} below.

Discrepancies
between $B_{c2}$ determined from resistive measurements and either
magnetisation \cite{welp} or specific heat \cite{carrington} data
have been reported, and it has been argued that the latter two better probe
the mean-field $B_{c2}$ than do resistivity measurements \cite{ando}. With the
definition of $B_{c2}$ used here for GHz measurements in the high field regime
the good agreement we find with d$B_{c2}$/d$T$ determined from low
field magnetisation measurements suggests that, in our case,
probing resistivity yields a $B_{c2}$ in accordance with the
mean-field value.

Although WHH theory includes paramagnetic and spin-orbit effects, in this
Letter we consider the model arising from orbital effects only, as applied by
Welp {\em et al} \cite{welp}. Whereas this model predicts a departure from the
slope $\alpha$ only at low $T$, we see clear evidence for a
departure below 74K. Previous measurements by
Nakagawa {\em et al} \cite{nakagawa}  provide convincing evidence for the
applicability of this model to YBCO for $B{\parallel}c$ (Fig.\ \ref{fig3}
inset). Near $T_c$ their data agree well with the
d$B_{c2}$/d$T$ slope determined previously \cite{welp} and application of the
WHH result:    \begin{equation}
B_{c2}(0)=0.7T_c(dB_{c2}/dT)|_{T_c} \label{eq1} \end{equation}
gives $B_{c2}^{\parallel}$(0)=112T, in close agreement with that measured.
Indeed, this agreement extends over the entire phase boundary (dotted line in
Fig.\ \ref{fig3} inset). A larger d$B_{c2}$/d$T$ for the case $B{\perp}c$, due
to anisotropy in coherence lengths ${\xi}_c$ out of plane and ${\xi}_{ab}$ in
plane, leads to a significantly larger $B_{c2}^{\perp}$(0)$>$600T in this
model. A deviation well below this value is clear in Fig.\ \ref{fig3}, however.
The good agreement between WHH and experiment for $B{\parallel}c$ and the
departure from expected behaviour for $B{\perp}c$ suggests that a different
mechanism may be responsible for the quenching of superconductivity in the
latter case. Misalignment of $B$ has been considered but cannot explain the
low $T$ results of Ref. \cite{dzurak} or the deviation observed here.

\begin{figure}
\vspace{-1.6cm}
\includegraphics[width=7cm,angle=270]{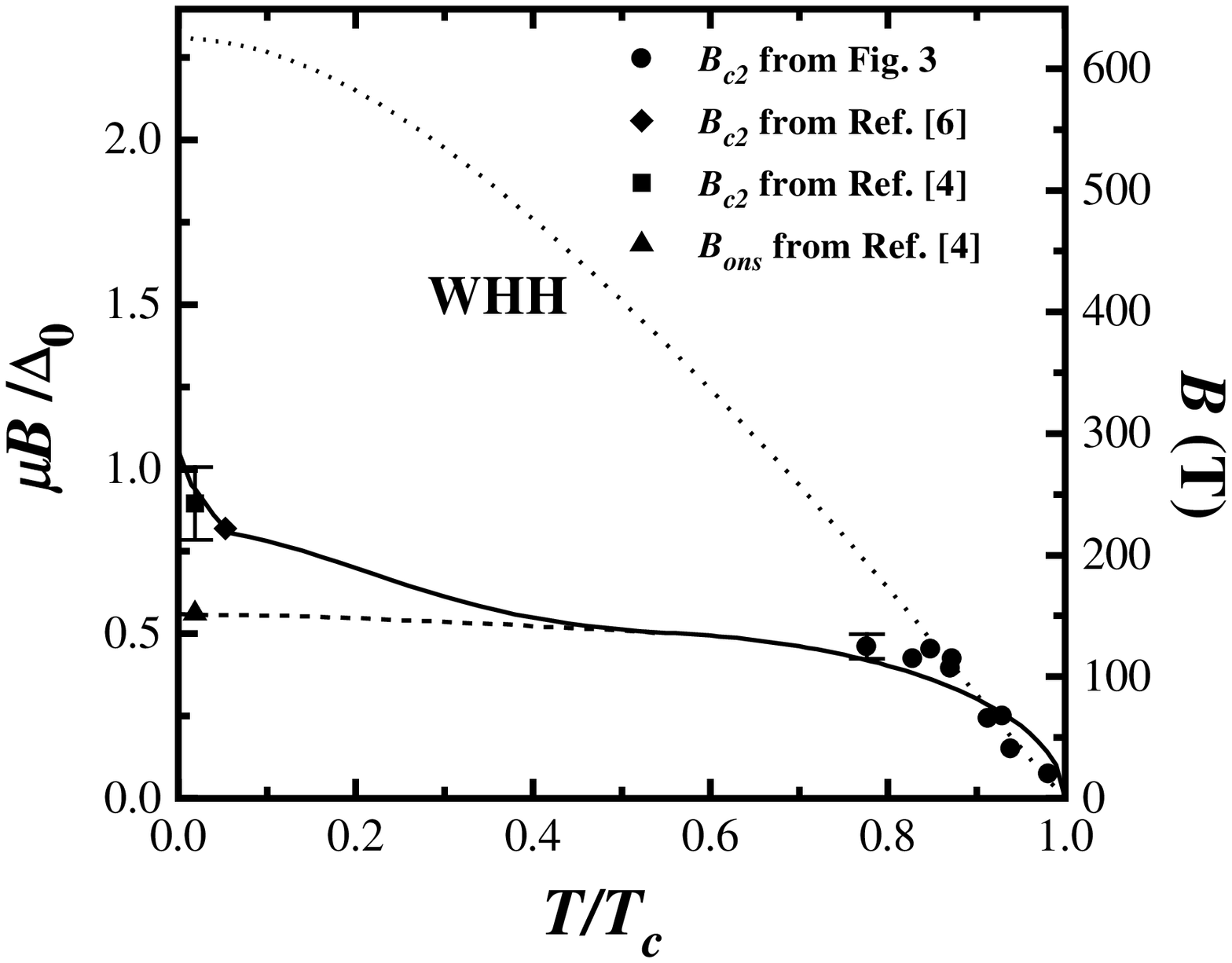}
\caption{Full phase diagram for YBCO for $B{\perp}c$. The $B_{c2}$ data from
Fig. 3 is replotted along with low $T$ data from Refs.
\protect\cite{dzurak,goettee}. The dotted line is the WHH phase
boundary \protect\cite{whh} calculated considering orbital effects only. The
solid line is the second order phase boundary, and the dashed line, the first
order BCS-FFLO phase boundary taken from Ref. \protect\cite{yang} which
considers only coupling between the spins and the applied $B$. The reduced
field is calculated with the energy gap defined as $\Delta_0$=2.14$T_c$
\protect\cite{yang}.}
\vspace{-.3cm}
\label{fig4} 
\end{figure}

Paramagnetic-limited upper critical fields have been observed in UPd$_2$Al$_3$
\cite{gloos} and $B_{c2}$$>$$B_p$ has recently been seen in (TMTSF)$_2$PF$_6$
\cite{lee}. The results in Ref. \cite{dzurak} provided the first evidence for
paramagnetic limiting in a high $T_c$ material. The paramagnetic
limit in YBCO is expected to occur at $B_p$${\sim}$170T, well above
$B_{c2}^{\parallel}$(0)=110T measured for $B{\parallel}c$ \cite{nakagawa}, and
well below $B_{c2}^{\perp}$(0)=640T predicted for $B{\perp}c$ with $T_c$=87K
(see Ref. \cite{welp}). The difference between WHH and experiment for
$B{\perp}c$ is shown clearly in Fig.\ \ref{fig4} where the WHH phase
boundary and the data from Fig.\ \ref{fig3} are plotted on a full phase
diagram. The departure from this phase
boundary is consistent with the results from flux compression measurements
\cite{dzurak,goettee} (also plotted) and provides further evidence for
paramagnetic limiting of $B_{c2}$ for $B{\perp}c$ \cite{note}. 

A possible fit to the data of Refs.
\cite{dzurak,goettee} for $B{\perp}c$ has
been obtained \cite{brooks} by including spin-orbit and paramagnetic parameters
in the WHH and Maki \cite{maki} models. Whilst it is instructive
to apply these models, the value of the parameter giving rise to
paramagnetic effects is tending to be unphysical \cite{brooks},
and spin-orbit scattering should be negligible in YBCO above a few Tesla since
it is in the clean limit \cite{yang}. Furthermore, the applicability of these
3D models to in-plane critical fields in layered superconductors is brought
into question by analysis \cite{klemm} which has found that there is a non-zero
temperature $T^*$$<$$T_c$ below which the normal cores of the vortices are
smaller than the lattice constant $d$, defined by
${\xi}_c(T^*)$=$d/\sqrt{2}=8.5$\AA. Below $T^*$ orbital effects should no
longer provide a mechanism for quenching of superconductivity and, in the
absence of paramagnetic and spin-orbit effects, $B_{c2}$ would be infinite.

A 3D-2D crossover is expected to occur near $T^*$ when
${\xi}_c$ becomes smaller than the inter-plane spacing.
${\xi}_c$(70K)${\sim}$8{\AA} is the separation between pairs of
CuO planes \cite{burns}, and ${\xi}_c$(80K)${\sim}$d=12{\AA} is the
lattice constant \cite{schneider}. The departure from the linear
behaviour of $B_{c2}(T)$ observed here at $T$=74K is almost midway between
these two characteristic temperatures. A theory which includes the
finite thickness of the superconducting layers in the cuprates, but neglects
paramagnetic effects \cite{schneider}, predicts a crossover in $B_{c2}(T)$ from
linear to non-linear behaviour at $T$=0.9$T_c$$\sim$78K, close to
the departure observed here. The fact that we see
$B_{c2}$ increase less rapidly with $T$ in the nonlinear regime, in contrast
to this theory \cite{schneider}, is most likely due to paramagnetic
limiting.

For
systems in which paramagnetic effects are important a
finite momentum or Fulde-Ferrel-Larkin-Ovchinnikov (FFLO) \cite{fflo}
superconducting state may exist. An extension of the original theory suggests
that a large ratio of orbital to paramagnetic terms,
${\beta}$=$\sqrt{2}B_{c2}(0)/B_p$, is favourable to the formation of such a
state, provided the superconductor is in the clean limit \cite{gruenberg}.
Here $B_{c2}(0)$ is that defined in Eq. 1. The first reported observation of
the FFLO state was in UPd$_2$Al$_3$ with ${\beta}$=2.4 \cite{gloos}. More
recently it has been suggested that the FFLO state should be enhanced in
quasi-2D superconductors \cite{shimahara}, making YBCO, which is in the clean
limit with ${\beta}$=5.7 for $B{\perp}c$, an ideal candidate. 

The phase
diagram predicted \cite{yang} for $d$-wave superconductivity in layered
materials with $B{\perp}c$ which accounts for coupling between the spins and
the applied $B$ includes a FFLO phase. A comparison with experiment is shown in
Fig.\ \ref{fig4} where the only free parameter is the $g$-factor, which is set
equal to 2. The agreement between this theory \cite{yang} and the low $T$ data
\cite{dzurak,goettee} is surprisingly good, supporting the conjecture
that quenching of superconductivity is driven by paramagnetic effects at low
$T$. In contrast, the data close to $T_c$ is better fitted by WHH, which is
probably due to the fact that the theory in Ref. \cite{yang} does not consider
orbital effects. We note that $B_{ons}$ for the data of Ref. \cite{dzurak}
coincides with the position of the first order phase transition between the
zero momentum and FFLO state. This may not be accidental since the
superfluid density and $J_c$ may be lower in the FFLO phase \cite{ys}.

In conclusion, we have found that for $B{\perp}c$
the upper critical field in YBCO follows the WHH phase boundary for
$T$$>$74K. A clear departure is observed below 74K near the
expected position of a 3D-2D crossover. The low $T$
data below the crossover is consistent with a theory \cite{yang} which
accounts for coupling of the spins to the applied $B$. This
suggests a transition from a high $T$ regime where superconductivity
is governed by orbital effects to a low $T$ regime where paramagnetic
effects dominate. Combined with the data of Refs. \cite{dzurak,goettee} this
constitutes strong evidence for the experimental realisation of
paramagnetic limiting in a high-$T_c$ cuprate superconductor. 

We thank K. Yang, S.L. Sondhi and R.H. McKenzie for detailed discussions and
comments on this manuscript, B. Sankrithyan, who
grew the film, and researchers at the Megagauss Laboratory, ISSP,
Tokyo University for expert technical help and advice.


\begin{references}
\bibitem{note} $B_{c2}$ was obtained in Ref. \cite{goettee} by taking the
precise zero crossing of the imaginary component of the saturated GHz signal,
guided by theoretical arguments. However the most significant feature in the
data defining the transition occurs at significantly lower $B$ and $B_{c2}$
obtained from this feature, as in Fig.\ \ref{fig2}, is plotted here.
\end{references}
\enddocument